\pdfoutput=1
\documentclass[aps, prd, superscriptaddress, notitlepage, 11pt, final, longbibliography]{revtex4-1}

\usepackage{graphicx}
\usepackage{color}
\usepackage{hyperref}

\usepackage{float}
\usepackage{amsmath, amssymb, mathrsfs}
\usepackage{subcaption}
\usepackage{epsf}
\usepackage{bbm}
\usepackage{comment}
\usepackage{cleveref}
\usepackage{comment}
\usepackage{natbib}
\usepackage{appendix}
\usepackage{here}

\renewcommand{\bm}[1]{{\mbox{\boldmath $#1$}}} 
\graphicspath{{./paper/}} 

\begin{document}
\title{A Modified Center-of-Mass Conservation Law in Finite-Domain Simulations of the Zakharov--Kuznetsov Equation}

\author{Nobuyuki Sawado~}
\email{sawadoph@rs.tus.ac.jp}
\affiliation{Department of Physics and Astronomy, Tokyo University of Science, Noda, Chiba 278-8510, Japan}

\author{Yuichiro Shimazaki~}
\email{shimazakitus@gmail.com}
\affiliation{Department of Physics and Astronomy, Tokyo University of Science, Noda, Chiba 278-8510, Japan}

\begin{abstract}
We investigate conservation laws of the two-dimensional Zakharov--Kuznetsov (ZK) equation, a natural higher-dimensional and non-integrable extension of the Korteweg--de Vries equation.
The ZK equation admits three scalar conserved quantities—mass, momentum, and energy—represented as $I_1$, $I_2$, and $I_3$, as well as a vector-valued quantity $\bm{I}_4$.
In high-accuracy numerical simulations on a finite double-periodic domain, most of these quantities are well preserved, while a systematic temporal drift is observed only in the $x$-component $I_{4x}$.
We show that the nontrivial evolution of $I_{4x}$ originates from an explicit boundary-flux contribution, which is induced by fluctuations of the solution and its spatial derivatives at the domain boundaries.
We successfully identify the source of the inaccuracy in the numerical solutions.
Motivated by this analysis, we define a modified center-of-mass quantity $I_{4x}^{\mathrm{mod}}$
and demonstrate its conservation numerically for single-pulse configurations.
The modified quantity thus provides a consistent conservation law for the ZK equation
and yields an appropriate description of center-of-mass motion in finite-domain numerical simulations.
\end{abstract}

\maketitle

\section{Introduction\label{sec:introduction}}

Conservation laws play a fundamental role in the analysis of nonlinear dispersive partial differential equations (PDEs).
They provide essential insight into the qualitative structure of solutions and form the theoretical basis for many numerical diagnostics.
In most of the derivations, the conserved quantities are obtained under an assumption that the solution and its sufficiently high-order spatial derivatives vanish at infinity.
In contrast, numerical solutions are obtained on a finite domain with periodic boundary conditions.
This approach is justified if the conservation laws still hold under the periodic truncation.
As we demonstrate in this work, this assumption is not universally justified: certain conserved quantities are intrinsically sensitive to boundary effects and cannot be faithfully represented by periodic boundary conditions on a finite domain.

In this context, it is natural to ask whether finite-domain simulations also satisfy conservation laws that are derived on the whole space.
To address this question, we consider the two-dimensional Zakharov--Kuznetsov (ZK) equation, a natural higher-dimensional generalization of the Korteweg--de Vries (KdV) equation.
While the ZK equation admits several conserved quantities on the whole space, their behavior under finite-domain truncation, especially for center-of-mass-type invariants, is far less understood.

For context, a classical and influential example of a nonlinear dispersive PDE is the KdV equation~\cite{Zabusky1965}.
It is a prototypical integrable system, possessing an infinite hierarchy of conservation laws that support stable soliton solutions~\cite{Miura1968}.
The structure of conserved quantities is fundamental both for understanding qualitative features of solutions and for constructing accurate and robust numerical schemes.
Moreover, conserved quantities underpin variational and stability theories for Hamiltonian dispersive systems (see, e.g.,~\cite{Zakharov2012}), and in the ZK case the linear stability of ground-state solitons has been rigorously linked to a Vakhitov--Kolokolov-type condition formulated via the conserved momentum~\cite{Kuznetsov2018}.

In contrast to fully integrable systems like the KdV equation, many multi-dimensional dispersive PDEs permit only a finite set of conservation laws and are often referred to as \emph{quasi-integrable} systems~\cite{Ferreira2013}.
The ZK equation belongs to this class.
It arises as an asymptotic model for ion--acoustic waves in magnetized plasma~\cite{Sagdeev1969, Zakharov1974} and has been extensively studied from both analytical and numerical perspectives~\cite{Kuznetsov1986, Iwasaki1990, Faminskii1995, deBouard1996, Munro1999, Linares2009, Johnson2010, Linares2011, Farah2012, Ribaud2012, Grunrock2013, Pilod2014, Ponce2015, Cote2016, Klein2021, Faminskii2022}.
The ZK equation admits three standard scalar conserved quantities—mass, momentum, and energy—denoted by $I_i$ $(i=1,2,3)$, as well as a vector-valued quantity $\bm{I}_4$ associated with the center of mass~\cite{Zakharov1974}.
The scalar invariants have been discussed extensively in the literature, and their numerical preservation has been examined in a variety of computational studies~\cite{Bridges2001, Nishiyama2012, Xu2012, Klein2023}.
In contrast, comparatively little attention has been paid to the vector-valued quantity that characterizes the motion of the center of mass.

The ZK equation possesses exact line-soliton solutions that can be regarded as higher-dimensional analogues of KdV solitons.
Such solutions do not decay in the $y$-direction and therefore take nonzero values at the $y$-boundaries on a finite computational domain.
To remain consistent with the analytical assumptions underlying conservation-law derivations, our numerical investigation focuses on fully localized pulse-type solutions that are isolated from both spatial boundaries~\cite{Iwasaki1990}.
This setting allows us to directly compare numerical measurements with the corresponding analytical conservation laws.

Since the explicit solutions are available only in special cases, numerical simulation plays a central role in the study of the ZK equation.
Fourier (pseudo-)spectral discretizations~\cite{Gottlieb1977, Trefethen2000}, often combined with exponential time-differencing Runge--Kutta schemes~\cite{Cox2002, Kassam2005}, have become standard tools due to their excellent accuracy and stability for dispersive equations~\cite{Chan1985, Gavete2008, Chen2011, Zavalani2014, Klein2014, Klein2021}.
These methods are known to reproduce the standard scalar conserved quantities with high accuracy~\cite{Bridges2001, Chen2011}.
However, the numerical behavior of the center-of-mass quantity, particularly its sensitivity to finite-domain truncation and boundary effects, has not been systematically investigated.

In this paper, we introduce a modified center-of-mass conserved quantity that remains well conserved in finite-domain simulations.
We perform high-precision numerical simulations of the two-dimensional ZK equation on a finite domain.
The standard scalar conserved quantities---mass, momentum, and energy---are preserved to high accuracy.
The $y$-component of the center-of-mass quantity is likewise well conserved.
In contrast, the $x$-component exhibits a clear and systematic temporal drift, despite being an exact conserved quantity of the equation.
We show that this deviation originates from boundary-flux contributions that arise in the finite-domain formulation.
The modified quantity is therefore defined by explicitly subtracting the analytically derived boundary contribution, and we demonstrate its conservation in numerical experiments.

The paper is organized as follows.
Section~\ref{sec:zk_equation} reviews the Zakharov--Kuznetsov equation and its conserved quantities.
Section~\ref{sec:modified} derives the finite-domain correction to the center-of-mass conservation law.
Section~\ref{sec:centroid} demonstrates the resulting uniform centroid motion for isolated pulse-type solutions.
Concluding remarks are given in Sec.~\ref{sec:summary}.

\section{Zakharov--Kuznetsov Equation\label{sec:zk_equation}}

The Zakharov--Kuznetsov (ZK) equation was originally introduced as a three-dimensional model for ion-acoustic waves in a plasma subject to a uniform magnetic field~\cite{Zakharov1974}.
It can be regarded as a natural multidimensional extension of the Korteweg--de Vries (KdV) equation and serves as a prototypical example of a quasi-integrable system~\cite{Ferreira2013}.
Although the term ``quasi-integrability'' is widely used, its precise mathematical meaning remains somewhat ambiguous.
A common working definition is that such systems possess only a finite number of exactly conserved quantities, in contrast to completely integrable systems, which admit an infinite hierarchy of conservation laws.
The ZK equation falls into this category, possessing three scalar invariants and one vector-valued invariant.
Despite this essential difference, the ZK equation retains several soliton-like features.

The ZK equation is given by
\begin{align}
\frac{\partial u}{\partial t}
+ \alpha\, u \frac{\partial u}{\partial x}
+ \frac{\partial}{\partial x} (\nabla^2 u) = 0,
\label{eq:zk_original}
\end{align}
where the Laplacian is defined as $\nabla^2 = \partial_x^2 + \partial_y^2$, and the parameter $\alpha$ controls the strength of the nonlinear term, commonly taken to be $\alpha=2$.
This parameter can be absorbed into the amplitude of the solution.
In fact, by introducing the rescaling $u=\beta v$, Eq.~\eqref{eq:zk_original} transforms into
\begin{align}
v_t + (\alpha\beta)\, v v_x + \partial_x \nabla^2 v = 0.
\nonumber
\end{align}
Thus, varying $\alpha$ corresponds to an amplitude rescaling together with associated changes in the characteristic length and time scales.
In our numerical experiments, we set $\alpha=12$.
This choice is motivated purely by numerical benefits.
First, a larger nonlinear coefficient produces solitary-wave profiles that are thinner and more compact for a given amplitude, thereby reducing boundary effects on a finite computational domain.
Second, it enhances nonlinear interactions, allowing collision dynamics to be observed over shorter evolution times, which is advantageous when exploring a broad range of parameters.
Note that the choice of $\alpha$ does not alter any fundamental mathematical properties of the ZK equation.

Equation~\eqref{eq:zk_original} admits metastable, isolated vortex-type structures exhibiting soliton-like behavior.
These solutions propagate with nearly constant velocity in a preferred direction.
To describe such waves, we consider traveling solutions propagating in the positive $x$-direction with speed $c$ and change to a system of coordinates $(\tilde{x},y)$ which moves with $c$ as
\begin{align}
\tilde{x} := x - ct,
\qquad
u(t,x,y) = U(\tilde{x}, y).
\nonumber
\end{align}
We rewrite eq.~\eqref{eq:zk_original} in the form
\begin{align}
-c\, U_{\tilde{x}}
+ \alpha\, U\, U_{\tilde{x}}
+ \partial_{\tilde{x}} \nabla^2 U = 0.
\nonumber
\end{align}
Integrating once with respect to $\tilde{x}$ and imposing decay at spatial infinity, we obtain
\begin{align}
\nabla^2 U = c\,U - \frac{\alpha}{2} U^{2}.
\label{eq:U_equation}
\end{align}
A steady progressive solitary-wave solution within our convention $\alpha=12$ is given by
\begin{align}
U_{\rm 1d} = \frac{c}{4}\,{\rm sech}^2\!
\left[\frac{\sqrt{c}}{2}(\tilde{x}\cos\theta+y\sin\theta)\right],
\nonumber
\end{align}
where $\theta$ denotes the inclination angle of the waveform.
It indicates that this solution is just a trivial embedding of the KdV soliton into two spatial dimensions.
On the other hand, \eqref{eq:U_equation} posseses another solution keeping circular 
symmetry~\cite{Iwasaki1990}.
To find it, we introduce cylindrical coordinates and rewrite \eqref{eq:U_equation} as
\begin{align}
\frac{1}{r}\frac{d}{dr}\left(r\frac{dU(r)}{dr}\right) = cU(r) - \frac{\alpha}{2}U(r)^2,
\label{eq:pulse}
\end{align}
where $r:=\sqrt{\tilde{x}^2+y^2}$.
We can obtain the solutions with the boundary condition $U\rightarrow0$ as $r\rightarrow\infty$ numerically.
The solution represents a spatially localized solitary pulse propagating in the positive $x$-direction with uniform speed $c$.

The solutions of Eqs.~\eqref{eq:U_equation} and \eqref{eq:pulse} exhibit soliton-like dynamics; however, unlike the KdV equation, their stability is not supported by an infinite sequence of conserved quantities.
Instead, the ZK equation admits only four conserved quantities~\cite{Zakharov1974}, which for $\alpha=12$ may be written as
\begin{align}
I_1 &:= \iint u \, dx\,dy
= \int i_1(y)\,dy,
\qquad
i_1(y):=\int u\,dx,
\label{eq:zkcq1}
\\[4pt]
I_2 &:= \frac{1}{2}\iint u^2\,dx\,dy,
\label{eq:zkcq2}
\\[4pt]
I_3 &:= \iint
\left(
\frac{1}{2}\,|\nabla u|^2 - 2u^3
\right) dx\,dy,
\label{eq:zkcq3}
\\[4pt]
\bm{I}_4 &:= \iint \bm{r}\,u\,dx\,dy
- 6\,t\,\bm{e}_x \iint u^2\,dx\,dy,
\label{eq:zkcq4}
\end{align}
where $\bm{r}=(x,y)$ denotes the spatial position vector and $\bm{e}_x$ is the unit vector in the $x$-direction.
Here, $I_1$ represents the mass, whose density $i_1(y)$ is conserved in a manner analogous to the KdV equation, reflecting a remnant of its one-dimensional structure.
The quantities $I_2$ and $I_3$ correspond to the momentum and energy, respectively, while the vector-valued invariant $\bm{I}_4$ is associated with the conservation law governing center-of-mass motion.
Since the ZK equation possesses only a finite number of invariants and lacks the infinite hierarchy characteristic of completely integrable systems, it is an archetypal example of a quasi-integrable system.

\section{Variation of the Center-of-Mass Conservation Law and Its Modification
\label{sec:modified}}

In this section, we demonstrate behavior of  the conserved quantities of the ZK equation in numerical simulations performed on a finite computational domain.
Formally, under either periodic or decaying boundary conditions on $\mathbb{R}^2$, the conserved quantities \eqref{eq:zkcq1}--\eqref{eq:zkcq4} should exactly be preserved.
In practical numerical computations, the problem must be solved on a finite domain, and then, the exact conservation properties are not guaranteed a priori.

In standard simulations on a rectangular finite domain of sufficiently large extent, the scalar invariants $I_1$, $I_2$, $I_3$, as well as the $y$-component $I_{4y}$ of the vector quantity $\bm{I}_4$, are conserved precisely when appropriate numerical methods are employed.
In contrast, the $x$-component $I_{4x}$ exhibits a systematic temporal drift when the solution is evolved on a doubly periodic domain.
This drift cannot be explained by numerical error or insufficient resolution; rather, it originates from the structural form of the conserved quantity itself.

On the infinite plane $\mathbb{R}^2$, all boundary contributions vanish as the solution and its derivatives decay at the infinity.
For the case of a finite computational domain, weak oscillatory components generated by the quasi-integrable dynamics of the ZK equation propagate to the boundaries.
As a result, even when the localized core of the solution remains well separated from the domain boundaries, the formal conservation of $I_{4x}$ is broken by the boundary-induced flux.

To elucidate this mechanism, we analyze a numerically generated localized traveling-pulse solution.
Since no exact analytical pulse-type solution of the two-dimensional ZK equation is known~\cite{Iwasaki1990}, we employ the numerical solitary pulse propagating in the positive $x$ direction.
We employ the parameter regimes where the pulse is sufficiently far from the domain boundaries.
In the following subsections, we examine the temporal variation of $I_{4x}$ and compare it with analytically derived boundary contributions.
Throughout the analysis, we find a modified center-of-mass quantity that restores the conservation law on a finite domain.
To exclude several trivial effects such as nonlinear interactions or pulse collisions,
we focus on the propagation of a single localized pulse, thereby isolating boundary-induced effects arising from weak radiative tails.

\subsection{Numerical Simulations}
We set a doubly periodic computational domain
$(x,y) \in [-L_x, L_x] \times [-L_y, L_y]$ with $(L_x,L_y)=(32,16)$.
The numbers of grid points are chosen as $(N_x,N_y)=(1024,512)$.
Time integration is performed over the interval $t \in [0,10]$ using a time step $\Delta t = 2.5\times10^{-4}$.
Spatial discretization is carried out using a Fourier pseudo-spectral method, combined with an exponential time-differencing fourth-order Runge--Kutta (ETDRK4) scheme for time integration.
These parameters are chosen to minimize numerical dissipation while ensuring adequate spatial resolution throughout the simulation.

\begin{figure}[htbp]
\centering
\begin{subfigure}{0.48\linewidth}
    \centering
    \includegraphics[width=\linewidth]{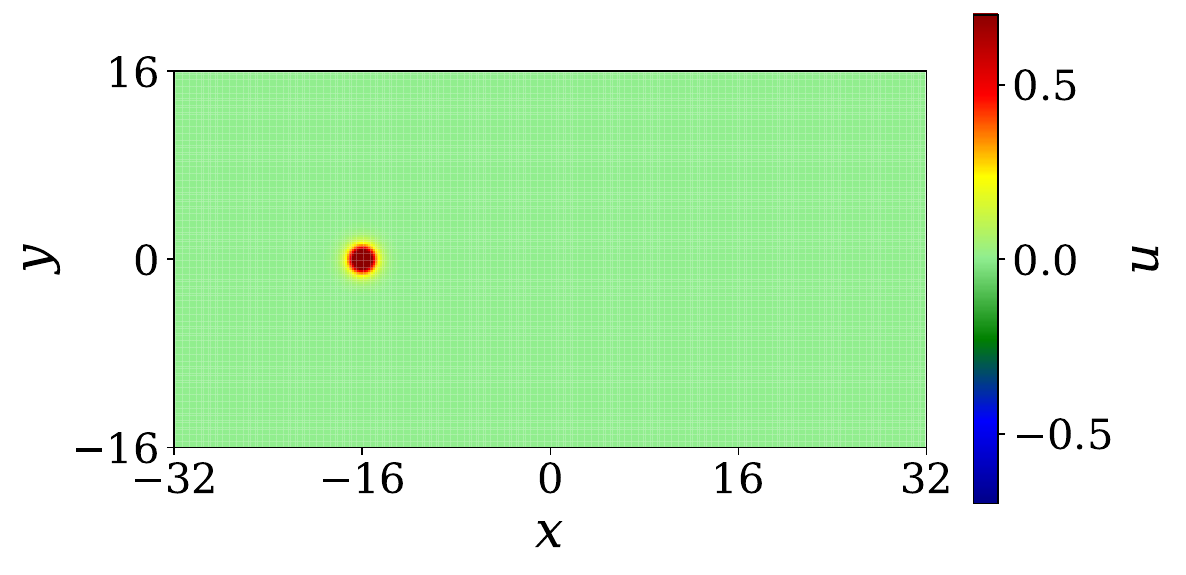}
\end{subfigure}
\hfill
\begin{subfigure}{0.48\linewidth}
    \centering
    \includegraphics[width=\linewidth]{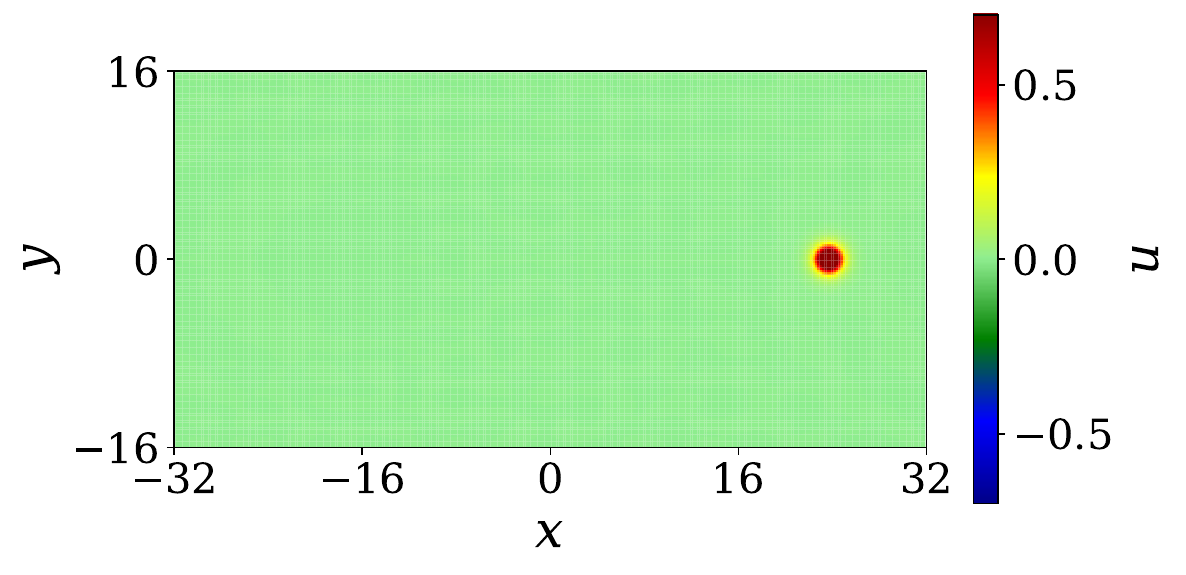}
\end{subfigure}
\caption{
Snapshots of a traveling solitary pulse of the ZK equation: the initial condition at $t=0$ (left) and the numerical solution at $t=10$ (right).
The pulse propagates stably without noticeable deformation.
Weak radiative tails, though barely visible at this scale, extend across the computational domain and generate nonzero boundary values.
}
\label{fig:snapshots_1soliton}
\end{figure}

Figure~\ref{fig:snapshots_1soliton} shows snapshots of the numerical solution at the initial time $t=0$ and the final time $t=10$.
The localized solitary pulse propagates with an approximately constant speed while maintaining its overall shape.
At the same time, small radiations emerge and spread throughout the computational domain.
Although these oscillatory tails are several orders of magnitude smaller than the amplitude of the pulse core, they induce nonzero boundary values of the solution and its spatial derivatives, which play a crucial role in behavior of the center-of-mass conservation law in finite-domain.

The standard scalar conserved quantities, namely the mass $I_1$, momentum $I_2$, and energy $I_3$, remain conserved throughout the simulation with relative deviations of order $\mathcal{O}(10^{-8})$ or smaller.
These small fluctuations are attributed to accumulated numerical errors from time integration and are negligible for the present discussion.

\begin{figure}[htbp]
\centering
\begin{subfigure}{0.48\textwidth}
\centering
\includegraphics[width=\linewidth]{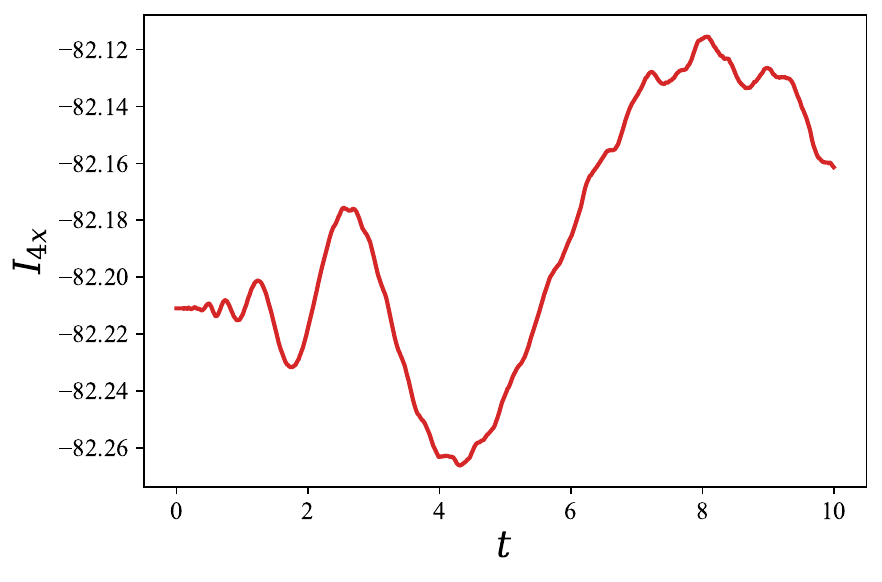}
\end{subfigure}
\hfill
\begin{subfigure}{0.48\textwidth}
\centering
\includegraphics[width=\linewidth]{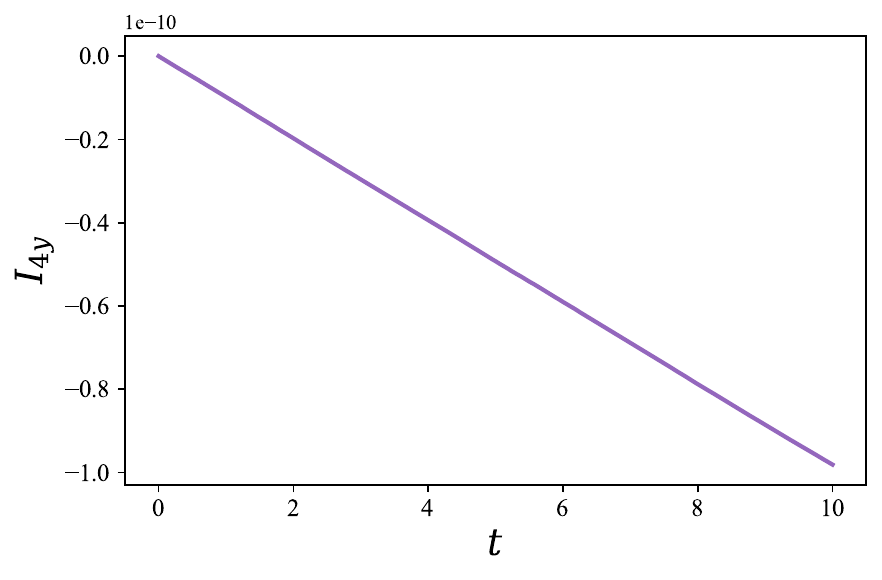}
\end{subfigure}
\caption{
Time evolution of the center-of-mass quantities: $x$-component $I_{4x}$ (left) and $y$-component $I_{4y}$ (right).
The $I_{4y}$ remains conserved to high accuracy, whereas the $I_{4x}$ exhibits a systematic drift induced by boundary flux contributions associated with weak radiative tails.
}
\label{fig:cqs_1soliton}
\end{figure}

Figure~\ref{fig:cqs_1soliton} displays the time evolution of the two components of the center-of-mass quantity $\bm{I}_4$.
The $y$-component $I_{4y}$ remains well conserved to numerical accuracy.
In contrast, the $x$-component $I_{4x}$ exhibits a clear and systematic temporal drift over the simulation interval.
Since the solitary pulse is always apart from the boundaries, the behavior is not a result of any structual change of the solution.
Instead, the drift originates from boundary-induced flux contributions generated by the weak radiative tails of the solution, as analyzed analytically in Appendix~\ref{appendix:i4x}.

\subsection{Analytical Study of the Boundary-Induced Variation in $I_{4x}$
\label{subsec:boundary_derivation}}
In this subsection, we analyze the boundary flux with respect to $I_{4x}$'s temporal drift.
We rewrite the ZK equation~\eqref{eq:zk_original} in a form suitable for identifying boundary contributions:
\begin{align}
u_t + \partial_x Q = 0,
\qquad Q := 6u^2 + \nabla^2 u.
\label{eq:zk_Q_form}
\end{align}
This representation makes explicit that the evolution of the center-of-mass quantity is governed by the flux of $Q$ through the domain boundaries.
We compute the time derivative of $I_{4x}$.
Since the analysis is a little complex, the process is described in Appendix A~\ref{appendix:i4x}.
\begin{align}
&\frac{dI_{4x}}{dt}
=
\mathcal{A}(t)+\mathcal{B}(t)+\mathcal{C}(t)+\mathcal{D}(t),
\label{eq:dI4}
\\ 
&
\mathcal{A}(t):=
- L_x \!\int_{-L_y}^{L_y}
\bigl[ Q(t, L_x, y) + Q(t, -L_x, y) \bigr] \, dy,
\label{eq:I4A}
\\
&\mathcal{B}(t):=
\int_{-L_y}^{L_y} \!\bigl[ u_x(t, L_x, y) - u_x(t, -L_x, y) \bigr] dy,
\label{eq:I4B}
\\
&\mathcal{C}(t):=
\int_{-L_x}^{L_x} \!\bigl[ u_y(t, x, L_y) - u_y(t, x, -L_y) \bigr] dx,
\label{eq:I4C}
\\
&
\mathcal{D}(t):=
\text{(other boundary terms with cubic or derivative form)}.
\label{eq:I4D}
\end{align}
Here, the individual boundary contributions are defined in Eqs.~\eqref{eq:I4A}--\eqref{eq:I4D}.
Due to the use of FFT-based differentiation, the numerical solution and spatial derivatives satisfy exact periodicity:
\begin{align}
u(t,\pm L_x,y) = u(t,\mp L_x,y), \quad
u_x(t,\pm L_x,y) = u_x(t,\mp L_x,y), \quad
u_y(t,x,\pm L_y) = u_y(t,x,\mp L_y),
\nonumber
\end{align}
and similarly for higher derivatives.
As a result, we easily confirm that $\mathcal{B}$, $\mathcal{C}$, and $\mathcal{D}$ become zero.
It is instructive to contrast this behavior with the infinite-domain case.
For solitary-pulse solutions, all fields and their derivatives decay exponentially as $|x|,|y|\to\infty$.
Consequently,
\begin{align}
L_x\, Q(t,\pm L_x,y) \longrightarrow 0
\qquad (L_x \to \infty),
\nonumber
\end{align}
and the boundary flux term $\mathcal{A}(t)$ vanishes since the terms $\mathcal{B}(t)$, $\mathcal{C}(t)$, and $\mathcal{D}(t)$ are identically zero,
so that $\bm{I}_4$ is conserved in the infinite-domain limit.

\begin{figure}[htbp]
\centering
\begin{subfigure}[t]{0.48\textwidth}
\centering
\includegraphics[width=\linewidth]{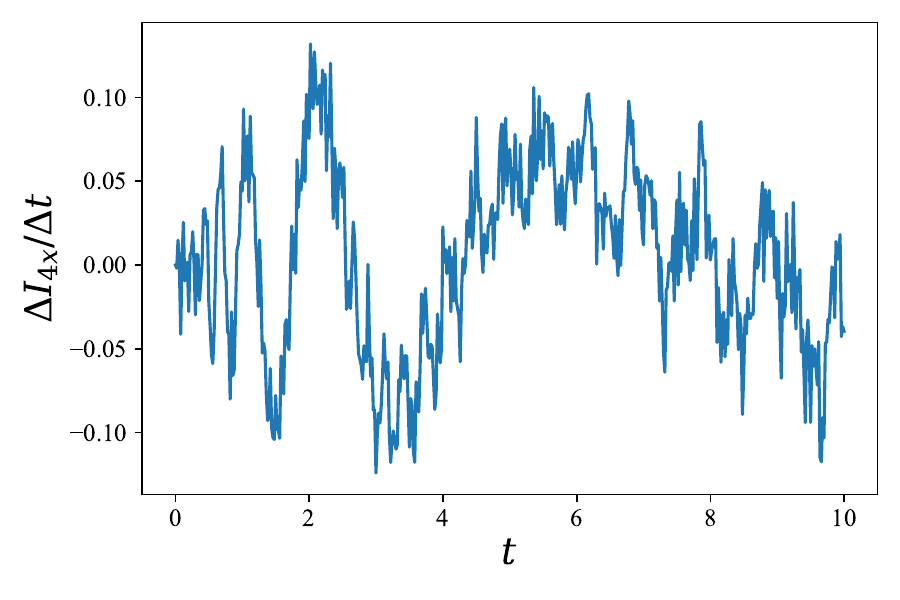}
\label{fig:di4xdt_num}
\end{subfigure}
\hfill
\begin{subfigure}[t]{0.48\textwidth}
\centering
\includegraphics[width=\linewidth]{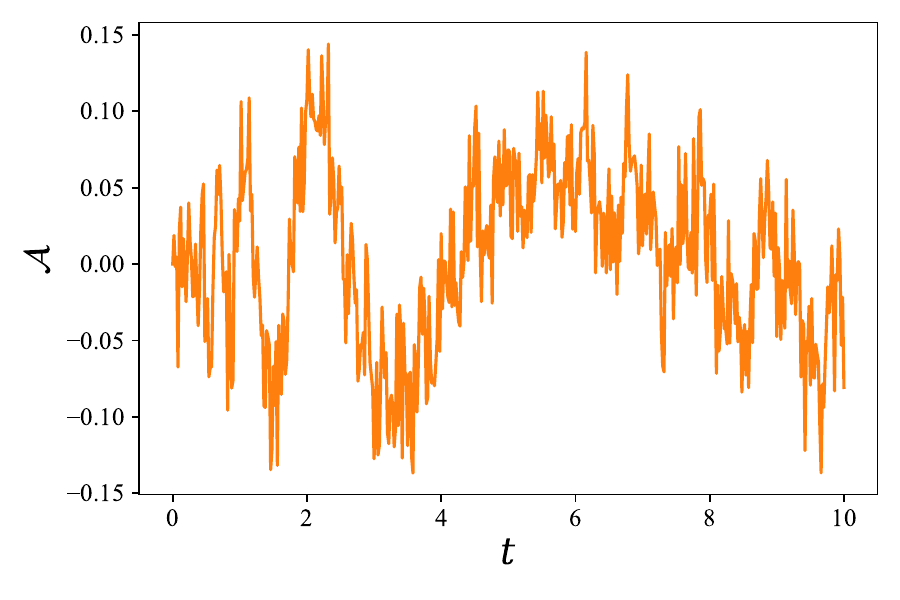}
\label{fig:di4xdt_termA}
\end{subfigure}
\vspace{-1.0cm}
\caption{
Comparison between the numerically evaluated time derivative $\Delta I_{4x}/\Delta t$ (left) and the analytically derived boundary flux contribution $\mathcal{A}(t)$ defined in Eq.~\eqref{eq:I4A} (right).
The two quantities exhibit the same temporal behavior over the entire simulation interval, confirming that the observed drift of $I_{4x}$ originates from the boundary flux contribution.
}
\label{fig:di4xdt_1soliton}
\end{figure}

\subsection{Modified Conservation Law for the Center of Mass}
The preceding subsection, we have demonstrated that the apparent drift of the center-of-mass-type quantity $I_{4x}(t)$ observed in finite-domain simulations originates entirely from the boundary flux term $\mathcal{A}(t)$.
A direct comparison of the absolute value of $I_{4x}(t)$ is not appropriate, since its magnitude depends on the initial position of the pulse and therefore does not carry intrinsic physical meaning.
In this subsection, we introduce a novel definition of the quantity that is comparable to the analysis.

To eliminate the boundary-induced drift in a systematic manner, we explicitly subtract the accumulated boundary contribution to define the modified quantity
\begin{align}
I_{4x}^{\mathrm{mod}}(t)
:= I_{4x}(t) - \int_{0}^{t} \mathcal{A}(\tau)\,d\tau .
\label{eq:modified_i4x}
\end{align}
From Eq.~\eqref{eq:dI4}, it immediately follows that
$\frac{d}{dt} I_{4x}^{\mathrm{mod}}(t) \approx 0$.

\begin{figure}[htbp]
\centering
\includegraphics[width=0.75\textwidth]{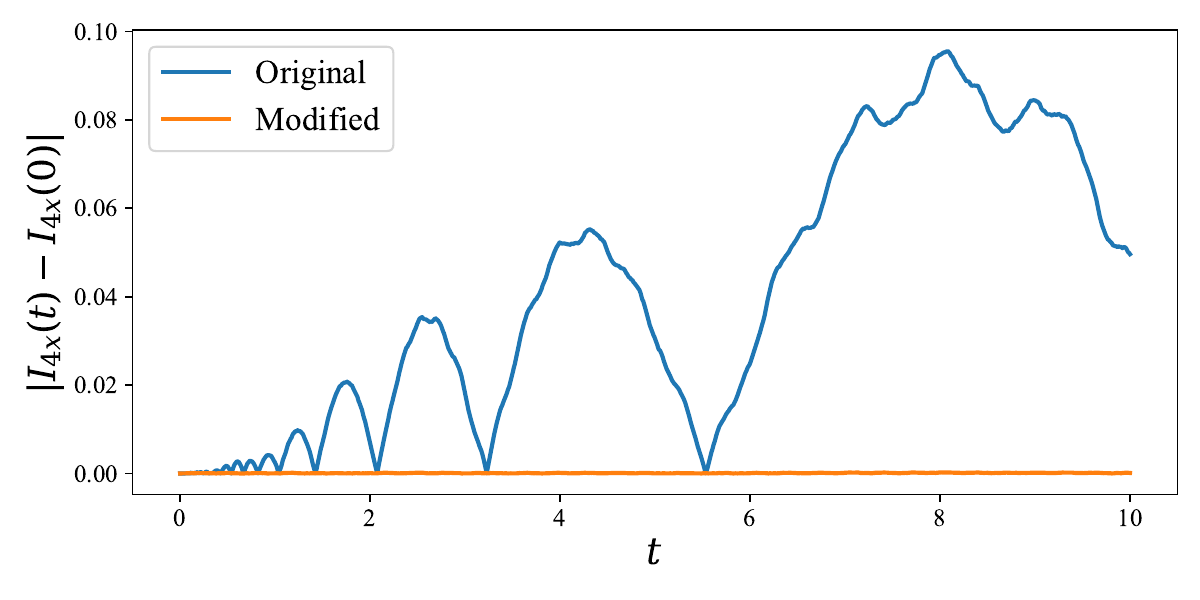}
\caption{
Time evolution of the deviation from the initial value for the center-of-mass-type quantity.
The solid line labeled ``Original'' represents $\lvert I_{4x}(t) - I_{4x}(0) \rvert$, while the line labeled ``Modified'' corresponds to
$\lvert I_{4x}^{\mathrm{mod}}(t) - I_{4x}(0) \rvert$.
}
\label{fig:modified_i4x_1soliton}
\end{figure}

Figure~\ref{fig:modified_i4x_1soliton} presents the time evolution of the deviations from the initial condition
$\lvert I_{4x}(t) - I_{4x}(0) \rvert$ and
$\lvert I_{4x}^{\mathrm{mod}}(t) - I_{4x}(0) \rvert$.
The subtraction of the initial condition removes trivial dependence on the position of the pulse and allows a direct comparison between the original and modified definitions.
While the original quantity exhibits a systematic growth due to boundary-induced flux, the modified quantity~\eqref{eq:modified_i4x} remains close to zero throughout the simulation.
The small residual fluctuations are attributable to numerical errors in evaluating the correction term and are negligible compared with the uncorrected drift.

\section{Analysis of centroid movement using conserved quantities\label{sec:centroid}}

In this section, we examine the physical implications of the modified center-of-mass conservation law.
Our primary focus is on the translational moving of a localized pulse, rather than on its detailed internal structure such as profile modulation or deformation.
We analyze the time evolution of the centroid expressed in terms of conserved quantities.
For localized, soliton-like solutions of dispersive nonlinear equations, the centroid provides a natural and physically meaningful measure of the effective position of a propagating structure.
When formulated through global invariants, the centroid dynamics are largely insensitive to fine-scale deformations and weak dispersive radiation.

For the ZK equation, the centroid motion in the $x$-direction can be expressed in terms of a specific combination of conserved quantities, namely the $x$-component of the vector invariant $\bm{I}_4$, the mass $I_1$, and the momentum $I_2$.
Together, these invariants determine the centroid position through integral properties of the solution.

As discussed in the previous section, boundary-induced fluxes affect the quantity $I_{4x}$ when the equation is solved on a finite double-periodic domain.
This effect leads to an artificial drift in the inferred centroid motion, which disturbs the intrinsic translational dynamics of the solution.
Therefore, to characterize the motion of the centroid accurately, we employ the modified conserved quantity.
We show that, with this correction, the centroid dynamics extracted from conserved quantities remain robust and physically meaningful even in finite-domain numerical simulations.

\subsection{Coordinate of the centroid}
We define the centroid (center-of-mass) position of a two-dimensional field $u(t,x,y)$ as
\begin{align}
\bm{r}_\textrm{c}(t)
:=
\frac{\iint \bm{r}\, u(t,x,y)\, dx\,dy}{\iint u(t,x,y)\, dx\,dy},
\label{eq:rc}
\end{align}
where $\bm{r}=(x,y)$.
This quantity can be expressed in terms of the conserved quantities introduced in Section~\ref{sec:zk_equation}.
Using the mass $I_1$~\eqref{eq:zkcq1}, the momentum $I_2$~\eqref{eq:zkcq2}, and the vector invariant $\bm{I}_4=(I_{4x},I_{4y})$~\eqref{eq:zkcq4}, the centroid coordinates are given by
\begin{align}
\bm{r}_\textrm{c}^{\mathrm{con}}(t)
=
\left(
\frac{I_{4x}(t) + 12 t\, I_2}{I_1},
\;
\frac{I_{4y}(t)}{I_1}
\right).
\label{eq:rc_def}
\end{align}

As shown in Section~\ref{sec:modified}, for the solutions of a finite double-periodic domain, $I_{4x}(t)$ is affected by boundary-induced fluxes.
The centroid coordinate then inferred from the original definition $\bm{r}_\textrm{c}^{\mathrm{con}}(t)$ exhibits a weak but systematic drift that is not related to the intrinsic translational dynamics of the localized pulse.
On the other hand, we define a modified centroid coordinate based on the modified center-of-mass invariant $I_{4x}^{\mathrm{mod}}$:
\begin{align}
x_\textrm{c}^{\mathrm{mod}}(t)
=
\frac{I_{4x}^{\mathrm{mod}}(t) + 12 t\, I_2}{I_1}
\label{eq:xc_mod}
\end{align}
successfully describes the proper centroid motion.
On the absolute scale, the centroid trajectories obtained from the original and modified definitions are nearly indistinguishable.
However, as demonstrated in the following subsection, the difference becomes clearly visible when examining the centroid velocity, which provides a more sensitive diagnostic of boundary-induced effects.

\subsection{Velocity of the centroid}
\begin{figure}[htbp]
\centering
\includegraphics[width=0.75\textwidth]{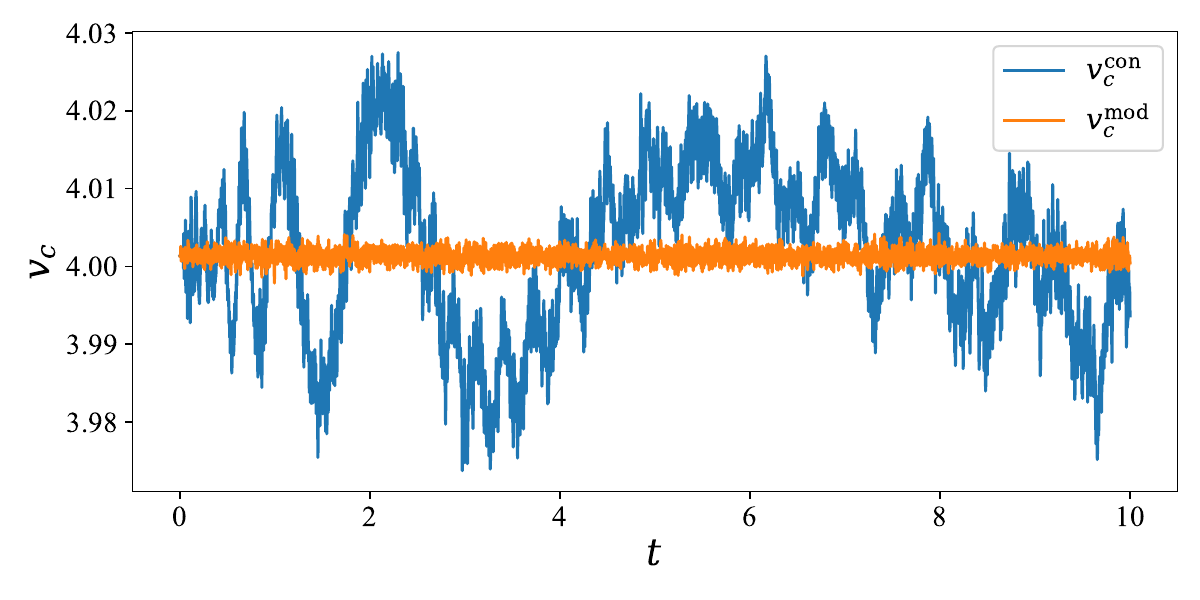}
\caption{
Comparison of centroid velocities computed from the original definition ($v_\textrm{c}^{\mathrm{con}}$, based on $I_{4x}$) and from the modified definition ($v_\textrm{c}^{\mathrm{mod}}$, based on $I_{4x}^{\mathrm{mod}}$).
Only the modified definition yields a nearly constant centroid velocity,
consistent with uniform translational motion of an isolated pulse.
}
\label{fig:centroid_v}
\end{figure}

A more sensitive indicator of the boundary-induced drift is obtained by considering the
centroid velocity, defined as the time derivative of the centroid coordinate,
\begin{align}
v_\textrm{c}(t)
:=
\frac{d x_\textrm{c}(t)}{dt}.
\end{align}
In practice, $v_\textrm{c}(t)$ is evaluated by finite differences applied to the discrete time series of $x_\textrm{c}(t)$ obtained from each centroid definition.
Figure~\ref{fig:centroid_v} compares the centroid velocities computed from $x_\textrm{c}^{\mathrm{con}}$ (based on the original invariant $I_{4x}$) and from $x_\textrm{c}^{\mathrm{mod}}$ (based on the modified invariant $I_{4x}^{\mathrm{mod}}$).
The velocity derived from the original definition exhibits noticeable temporal fluctuations,
reflecting the accumulation of boundary-induced drift already identified at the level of the conserved quantity.
On the other hand, the velocity obtained from the modified definition remains nearly constant over the entire simulation interval.
Apart from small high-frequency oscillations attributable to the finite-difference approximation, no systematic temporal variation is observed, indicating uniform translational motion of the centroid.

This observation is further supported by simple statistical measures.
For the velocity obtained from $x_\textrm{c}^{\mathrm{con}}$, the variance is
$\mathrm{var}(v_\textrm{c}) = 2.43\times10^{-4}$ with a standard deviation
$\mathrm{std}(v_\textrm{c}) = 1.56\times10^{-2}$.
In contrast, the velocity computed from $x_\textrm{c}^{\mathrm{mod}}$ exhibits a significantly reduced variance,
$\mathrm{var}(v_\textrm{c}) = 1.93\times10^{-6}$ and
$\mathrm{std}(v_\textrm{c}) = 1.39\times10^{-3}$,
corresponding to a reduction by more than two orders of magnitude.
Although the mean velocities in both cases agree closely, only the modified definition can realize nearly constant centroid velocity.

These results demonstrate that boundary-induced variations of $I_{4x}$ can significantly affect the extraction of physically relevant quantities, even when their impact on the centroid position itself appears negligible.
By compensating for this effect, the modified conservation law yields a centroid velocity that accurately reflects the intrinsic translational dynamics of the localized pulse.
The modification of \eqref{eq:modified_i4x} is not merely a minor correction, but is essential for obtaining a physically consistent description of centroid motion in finite-domain numerical simulations.
This clearly indicates that boundary-induced effects can contaminate dynamical diagnostics, even when their influence on integral quantities or centroid positions appears visually insignificant.

Although the present numerical study employs a localized pulse solution of the ZK equation as a representative example, the underlying mechanism is not specific to pulse-type structures nor to the ZK equation.
It seems to be a quite general property of several nonlinear dispersive equations based on numerical simulation: when global invariants are evaluated on finite computational domains, boundary-induced fluxes can subtly but systematically distort derived physical quantities.

To highlight this broader relevance, Appendix~\ref{appendix:kdv} examines an analogous situation for the integrable Korteweg--de Vries (KdV) equation.
Even in this integrable setting, ad hoc initial conditions tend to generate weak dispersive radiation, which in turn induces a boundary-related variation in the center-of-mass invariant.
A modified conservation law can again be constructed to restore the correct invariant behavior, demonstrating that the present correction principle applies beyond the quasi-integrable ZK framework.
This suggests that modified conserved quantities provide a useful and broadly applicable tool for extracting physically meaningful dynamics from finite-domain numerical simulations of dispersive systems.

\section{Summary\label{sec:summary}}

We have performed high-accuracy numerical simulations of the two-dimensional Zakharov--Kuznetsov (ZK) equation, a quasi-integrable nonlinear dispersive partial differential equation that admits several global invariants.
In particular, the ZK equation possesses three scalar conserved quantities—the mass $I_1$, momentum $I_2$, and energy $I_3$—as well as a vector-valued quantity $\bm{I}_4=(I_{4x}, I_{4y})$ associated with the center-of-mass dynamics.
Our simulations confirm that $I_1$, $I_2$, $I_3$, and the $y$-component $I_{4y}$ are conserved to high accuracy throughout the computation.
In contrast, the $x$-component $I_{4x}$ exhibits a systematic temporal drift, despite being formally classified as a conserved quantity.

We demonstrated that this apparent non-conservation originates from boundary-induced flux terms associated with weak radiative components of the solution.
A direct quantitative comparison shows that the analytically derived boundary flux precisely accounts for the numerically observed variation of $I_{4x}$.
This establishes that the drift is an intrinsic finite-domain effect rather than a consequence of numerical inaccuracies or deficiencies in the time-integration scheme.

For this issue, we introduced a modified conserved quantity for the center of mass, denoted by $I_{4x}^{\mathrm{mod}}$, which explicitly subtracts the accumulated boundary flux contribution.
For the single-pulse configurations considered in this study, numerical experiments demonstrate that the corrected quantity remains nearly constant and closely tracks its initial value over the entire simulation interval.
This confirms that the boundary-induced drift of $I_{4x}$ can be removed in a controlled and quantitatively consistent manner.

The physical significance of the modified conservation law was further elucidated through an analysis of centroid dynamics.
While centroid positions and velocities inferred from the original definition of $I_{4x}$ exhibit spurious drift and temporal fluctuations, those obtained from $I_{4x}^{\mathrm{mod}}$ display uniform translational motion, consistent with the expected kinematics of an isolated traveling structure.
This clearly demonstrates that the proposed modification is essential for extracting physically meaningful centroid dynamics from finite-domain numerical simulations.

Moreover, conservation-law-based diagnostics have also played an important role in stability considerations (e.g., \cite{Kuznetsov2018}).
In this context, the (modified) conserved quantity for center of mass, $\bm{I}_4$, provides a complementary and practically useful diagnostic: it offers a robust way to separate genuine shape deformation from neutral translational dynamics in finite-domain simulations.
This feature is particularly valuable in multi-pulse interactions, where accurate centroid tracking facilitates a systematic assessment of collision-induced displacement and phase shifts, as well as weak radiative effects, even when individual pulses partially overlap.

From the above considerations, we found that conservation-law-based diagnostics is a powerful tool for characterizing the motion of localized structures in most nonlinear dispersive systems.
At the same time, they highlight the crucial role of boundary-induced fluxes when such equations are studied on finite computational domains.
The modified conservation-law framework developed here offers a systematic and broadly applicable approach for extracting essential physical property from numerical data with the spurious boundary effects.
Our approach is expected to be relevant to a wide class of dispersive equations admitting solitons, soliton-like pulses, or other coherent structures.

\vspace{0.5cm}
\noindent {\bf Acknowledgments}

The authors thank Luiz Agostinho Ferreira, Atsushi Nakamula, Ryu Sasaki, Kiori Obuse and Kouichi Toda 
for fruitful discussions and valuable comments.
The authors are grateful to E. A. Kuznetsov for helpful comments and suggestions.
N. S.  is supported in part by JSPS KAKENHI Grant Number JP23K02794.

\begin{appendix}
\section{Boundary Effects on \texorpdfstring{$I_{4x}$}{I_{4x}}\label{appendix:i4x}}

We briefly investigate the boundary contributions responsible for the temporal variation of the quantity $I_{4x}$ defined in~\eqref{eq:zkcq4} when the Zakharov--Kuznetsov equation is solved on a finite domain.
The derivation is included for completeness and to clarify the origin of the boundary-induced flux terms discussed in the main text.

For convenience, we introduce the auxiliary quantities
\begin{align}
X(t) := \iint_D x\, u(t,x,y)\, dxdy,
\quad
P(t) := \iint_D u(t,x,y)^2\, dxdy,
\quad
(x,y)\in D:[-L_x,L_x]\times[-L_y,L_y],
\end{align}
in terms of which $I_{4x}$ can be written as
\begin{align}
I_{4x}(t) = X(t) - 6t\,P(t).
\label{eq:I4x_def_app}
\end{align}

We first compute the time derivative of $X(t)$.
Using the evolution equation $u_t + \partial_x Q = 0$ with $Q = 6u^2 + \Delta u$, we obtain
\begin{align}
\dot{X}(t)
&= \iint_D x\,u_t\,dxdy
= -\iint_D x\,\partial_x Q\,dxdy
\nonumber \\
&= - \int_{-L_y}^{L_y} [\, x Q(t,x,y) \,]_{x=-L_x}^{x=+L_x} dy
+ \iint_D Q(t,x,y)\,dxdy .
\label{eq:Xdot1_app}
\end{align}
The volume integral of $Q$ over $D$ is decomposed as
\begin{align}
\iint_D Q\,dxdy
&= \iint_D \nabla^2 u\,dxdy + 6 \iint_D u^2\,dxdy,
\nonumber \\
\iint_D \nabla^2 u\,dxdy
&= \int_{-L_y}^{L_y} [\, u_x(t,L_x,y) - u_x(t,-L_x,y) \,] dy
\nonumber \\
&\quad + \int_{-L_x}^{L_x} [\, u_y(t,x,L_y) - u_y(t,x,-L_y) \,] dx .
\end{align}
Substituting these expressions into~\eqref{eq:Xdot1_app}, we obtain
\begin{align}
\dot{X}(t)
&= - \int_{-L_y}^{L_y} \!\bigl( L_x Q(t,L_x,y) - (-L_x) Q(t,-L_x,y) \bigr) dy
\nonumber \\
&\quad + \int_{-L_y}^{L_y} [\, u_x(t,L_x,y) - u_x(t,-L_x,y) \,] dy
\nonumber \\
&\quad + \int_{-L_x}^{L_x} [\, u_y(t,x,L_y) - u_y(t,x,-L_y) \,] dx
+ 6 P(t).
\label{eq:Xdot_final_app}
\end{align}

Next, we evaluate the time derivative of $P(t)$:
\begin{align}
\dot{P}(t)
&= 2 \iint_D u\,u_t\,dxdy
= -2 \iint_D u\,\partial_x Q\,dxdy
\nonumber \\
&= -2 \int_{-L_y}^{L_y} [\, u Q \,]_{x=-L_x}^{x=+L_x} dy
+ 2 \iint_D u_x Q\,dxdy .
\label{eq:Pdot2_app}
\end{align}
The remaining volume integral can be expanded as
\begin{align}
2 \iint_D u_x Q\,dxdy
&= 2 \iint_D (u_x u_{xx} + u_x u_{yy})\,dxdy
+ 12 \iint_D u_x u^2\,dxdy,
\label{eq:Pdot_expand_app}
\end{align}
with individual terms evaluated as
\begin{align}
2 \iint_D u_x u_{xx}\,dxdy
&= \int_{-L_y}^{L_y} [\, u_x(t,L_x,y)^2 - u_x(t,-L_x,y)^2 \,] dy,
\label{eq:Pdot_i_app}
\\
2 \iint_D u_x u_{yy}\,dxdy
&= 2 \int_{-L_x}^{L_x} [\, u_x u_y \,]_{y=-L_y}^{y=+L_y} dx
- \int_{-L_y}^{L_y} [\, u_y^2 \,]_{x=-L_x}^{x=+L_x} dy,
\label{eq:Pdot_ii_app}
\\
12 \iint_D u_x u^2\,dxdy
&= 4 \int_{-L_y}^{L_y} [\, u^3 \,]_{x=-L_x}^{x=+L_x} dy.
\label{eq:Pdot_iii_app}
\end{align}
Combining these results yields
\begin{align}
\dot{P}(t)
&= \int_{-L_y}^{L_y}
\Bigl[\, -2uQ + u_x^2 - u_y^2 + 4u^3 \,\Bigr]_{x=-L_x}^{x=+L_x} dy
\nonumber \\
&\quad + 2 \int_{-L_x}^{L_x} [\, u_x u_y \,]_{y=-L_y}^{y=+L_y} dx .
\label{eq:Pdot_final_app}
\end{align}

Finally, using~\eqref{eq:Xdot_final_app} and~\eqref{eq:Pdot_final_app}, the time derivative of $I_{4x}$ is obtained as
\begin{align}
\frac{d}{dt} I_{4x}
= \dot{X}(t) - 6P(t) - 6t\,\dot{P}(t),
\end{align}
which consists entirely of boundary contributions.
This explicit expression confirms that, on a finite domain, the quantity $I_{4x}$ is generally subject to boundary-induced fluxes, even though it is formally conserved in the infinite-domain setting.

\vspace{0.5cm}

\section{Study of a One-Dimensional Integrable System: Korteweg--de Vries Equation}
\label{appendix:kdv}

We demonstrate that a center-of-mass-type conserved quantity also exists for a representative $(1+1)$-dimensional nonlinear dispersive system, namely the integrable Korteweg--de Vries (KdV) equation.
This example serves to illustrate that the boundary-induced effects discussed in the main text are not specific to the Zakharov--Kuznetsov equation, but arise more generally in finite-domain simulations of dispersive systems.

We consider the KdV equation in the form
\begin{align}
u_t + 6\,u\,u_x + u_{xxx} = 0,
\label{eq:kdv_app}
\end{align}
where $u(t,x)$ is a real-valued scalar field.
All computations are performed on a finite periodic domain $x \in [-L_x, L_x]$.

For typical traveling-wave solutions of~\eqref{eq:kdv_app}, there exists a conserved quantity associated with the center of mass, given by
\begin{align}
I_4(t)
=
\int_{-L_x}^{L_x} x\,u(t,x)\,dx
-
3t \int_{-L_x}^{L_x} u(t,x)^2\,dx .
\label{eq:kdv_I4_app}
\end{align}
This quantity is the one-dimensional analogue of the $x$-component of the center-of-mass invariant $\bm{I}_4$ introduced for the ZK equation in the main text.

When the initial condition is chosen as an exact soliton solution of~\eqref{eq:kdv_app}, the numerical solution propagates rigidly at constant speed without deformation.
In this case, the quantity $I_4$ is preserved up to numerical accuracy, and finite-domain effects remain negligible.

\begin{figure}[htbp]
\centering
\begin{subfigure}{0.31\textwidth}
\centering
\includegraphics[width=\linewidth]{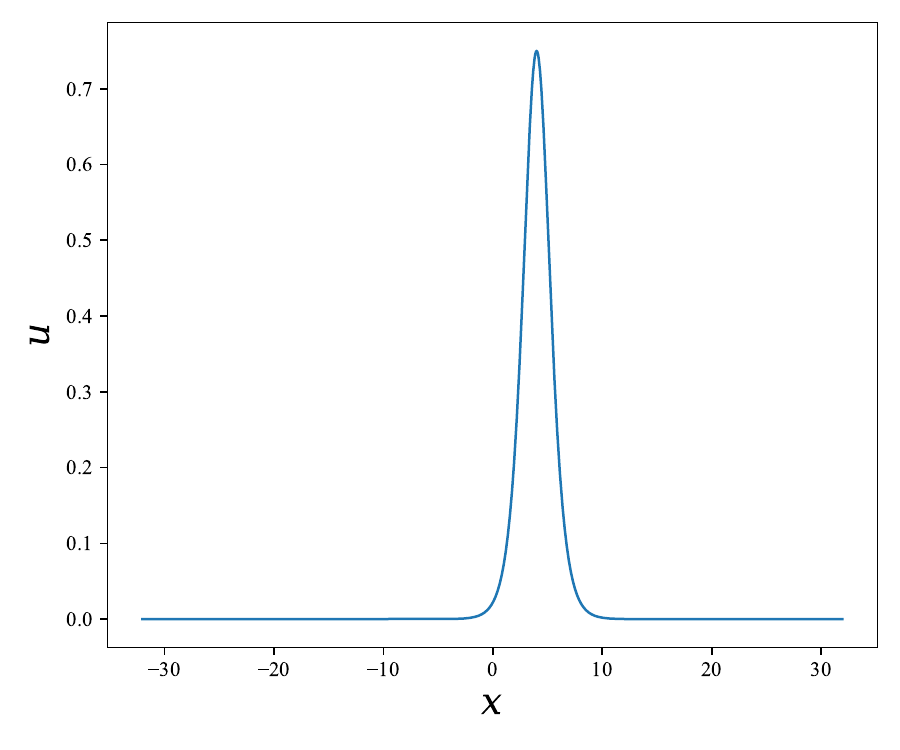}
\end{subfigure}
\begin{subfigure}{0.31\textwidth}
\centering
\includegraphics[width=\linewidth]{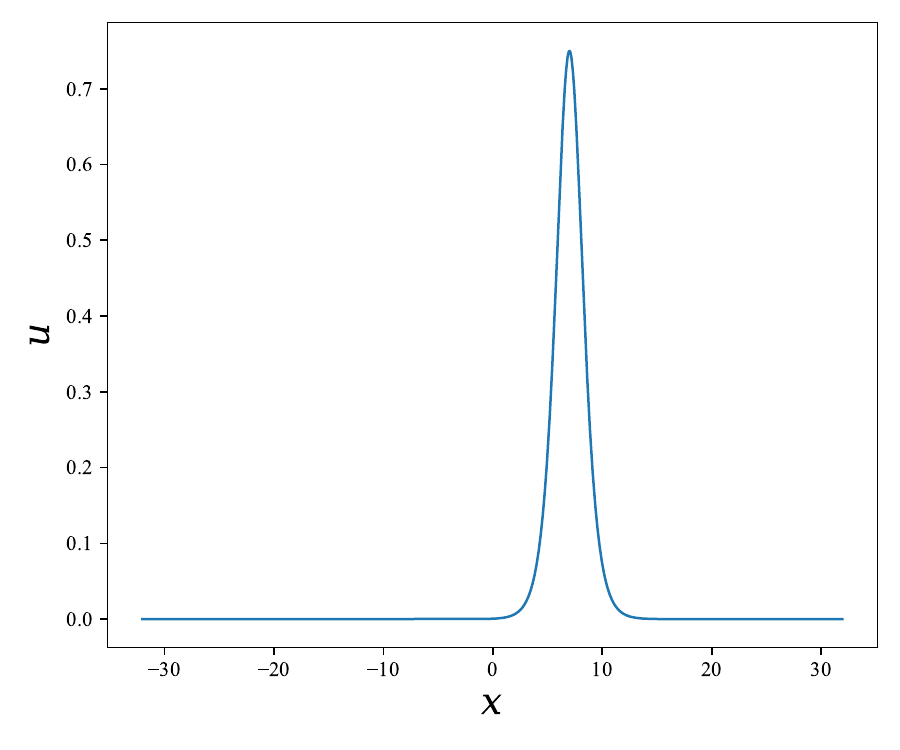}
\end{subfigure}
\begin{subfigure}{0.31\textwidth}
\centering
\includegraphics[width=\linewidth]{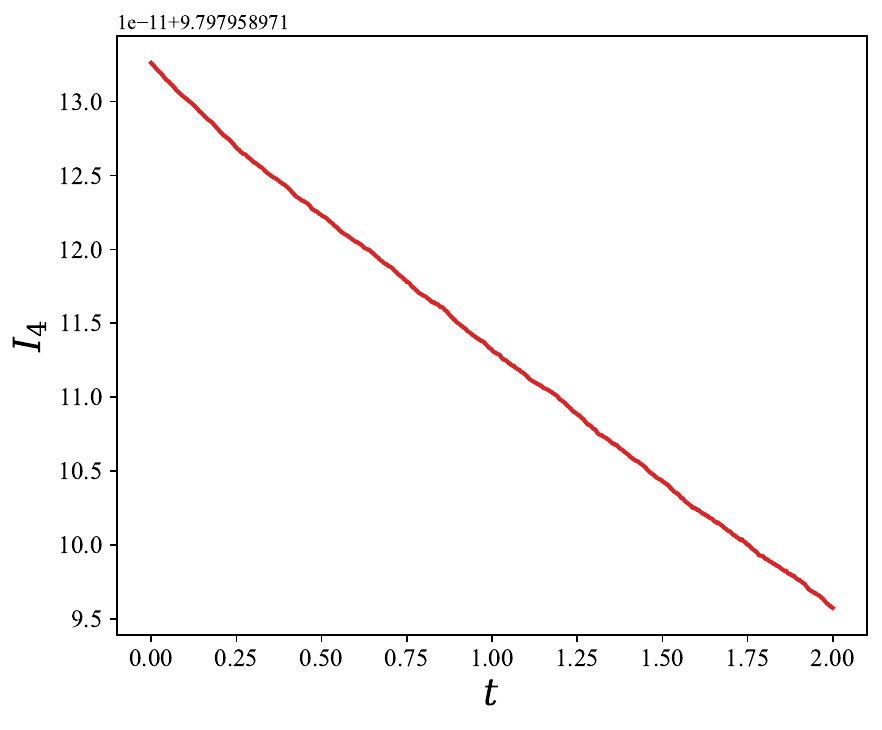}
\end{subfigure}
\begin{subfigure}{0.31\textwidth}
\centering
\includegraphics[width=\linewidth]{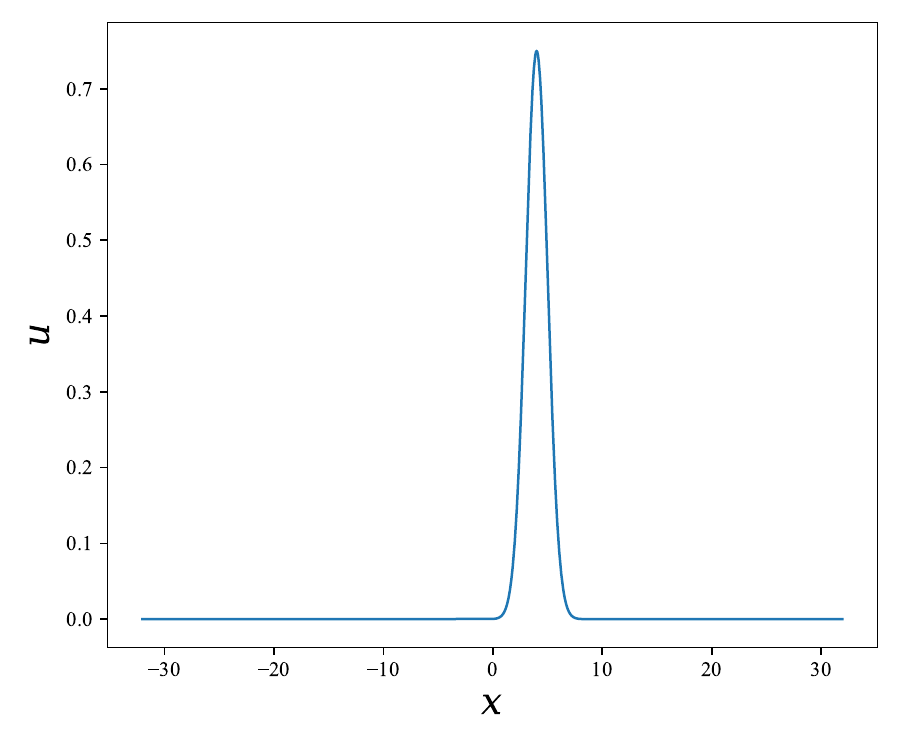}
\end{subfigure}
\begin{subfigure}{0.31\textwidth}
\centering
\includegraphics[width=\linewidth]{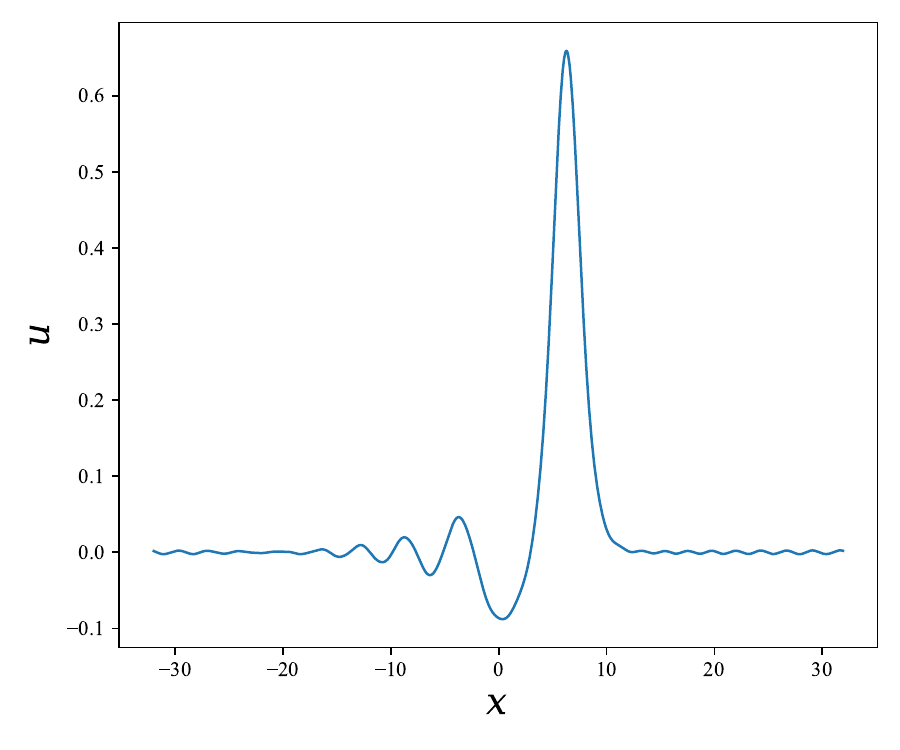}
\end{subfigure}
\begin{subfigure}{0.31\textwidth}
\centering
\includegraphics[width=\linewidth]{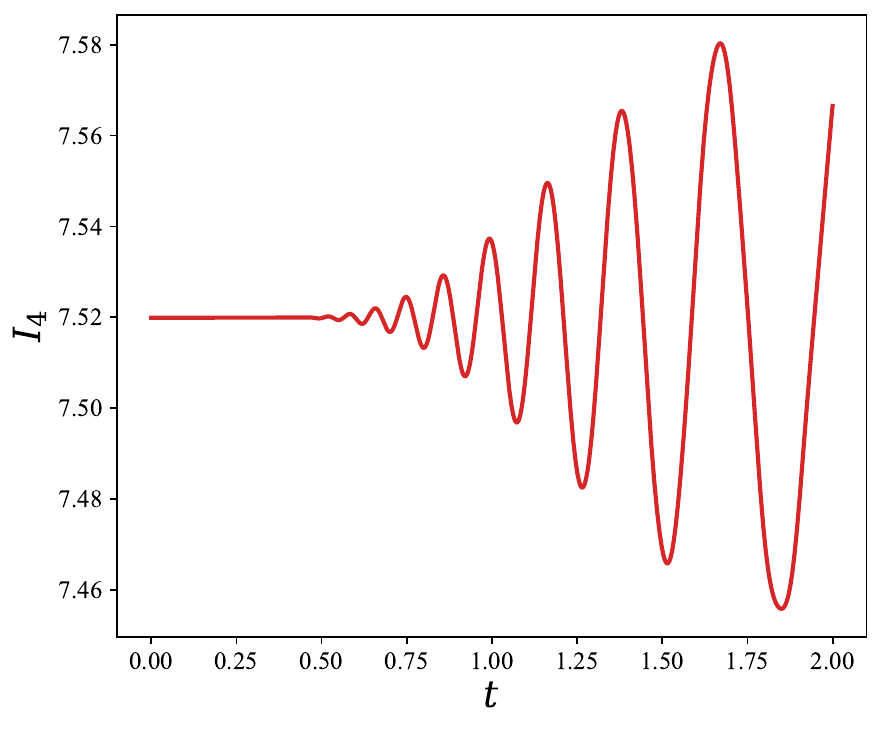}
\end{subfigure}
\caption{
Comparison of soliton (top row) and Gaussian (bottom row) initial conditions for the KdV equation.
Left: Initial profiles at $t=0$.
Middle: Evolved profiles at $t=2$.
Right: Time evolution of $I_4$.
The soliton propagates rigidly and preserves the $I_4$, whereas the Gaussian initial condition generates dispersive radiation and exhibits boundary-induced drift in $I_4$.
}
\label{fig:kdv_profiles}
\end{figure}

In contrast, when more generic initial conditions are employed, such as Gaussian profiles that are not exact soliton solutions, the solution develops weak dispersive radiation.
This radiation separates from the main pulse and eventually reaches the computational boundaries, leading to an apparent violation of the conservation of $I_4$.
Figure~\ref{fig:kdv_profiles} compares the solution profiles and the corresponding behavior of $I_4$ for soliton and Gaussian initial conditions.
While the soliton preserves both its shape and the conserved quantity, the Gaussian initial condition generates visible dispersive ripples and exhibits a systematic temporal drift in $I_4$.

Following the same strategy as employed for the ZK equation, the time derivative of $I_4$ in a finite domain can be expressed entirely in terms of boundary flux contributions.
We therefore define a modified center-of-mass quantity by explicitly subtracting the accumulated boundary flux,
\begin{align}
I_4^{\mathrm{mod}}(t)
=
I_4(t)
-
\int_0^t \mathcal{B}(\tau)\,d\tau,
\label{eq:kdv_I4_mod_app}
\end{align}
where $\mathcal{B}$ denotes the boundary flux evaluated at $x=\pm L_x$.
By construction, the modified quantity $I_4^{\mathrm{mod}}$ remains conserved even in the presence of dispersive radiation.

\begin{figure}[htbp]
\centering
\begin{subfigure}{0.45\textwidth}
\centering
\includegraphics[width=\linewidth]{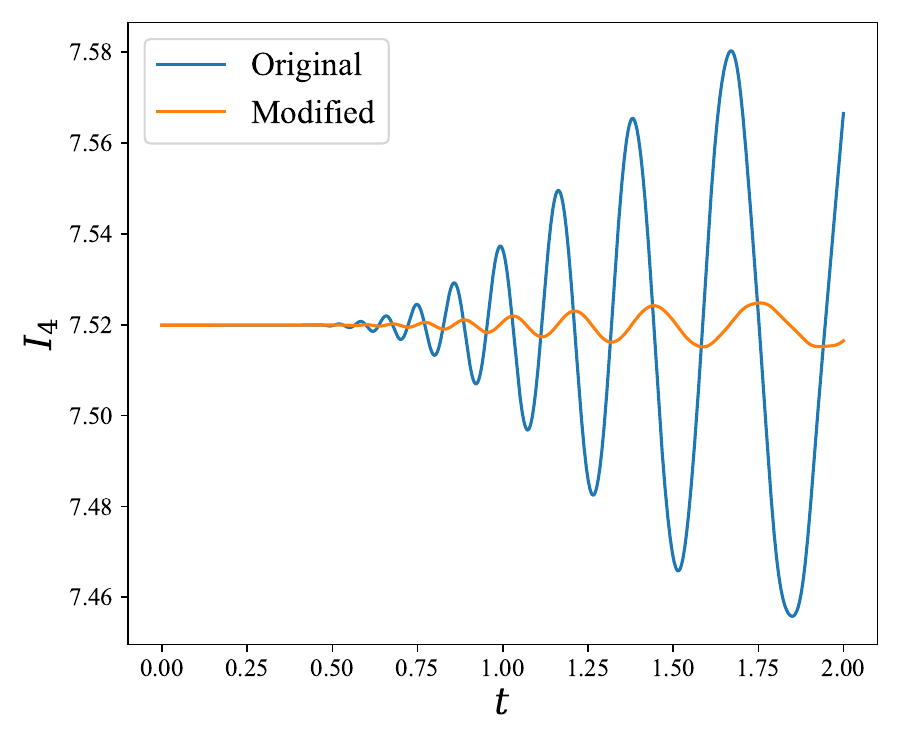}
\end{subfigure}
\hspace{1em}
\begin{subfigure}{0.45\textwidth}
\centering
\includegraphics[width=\linewidth]{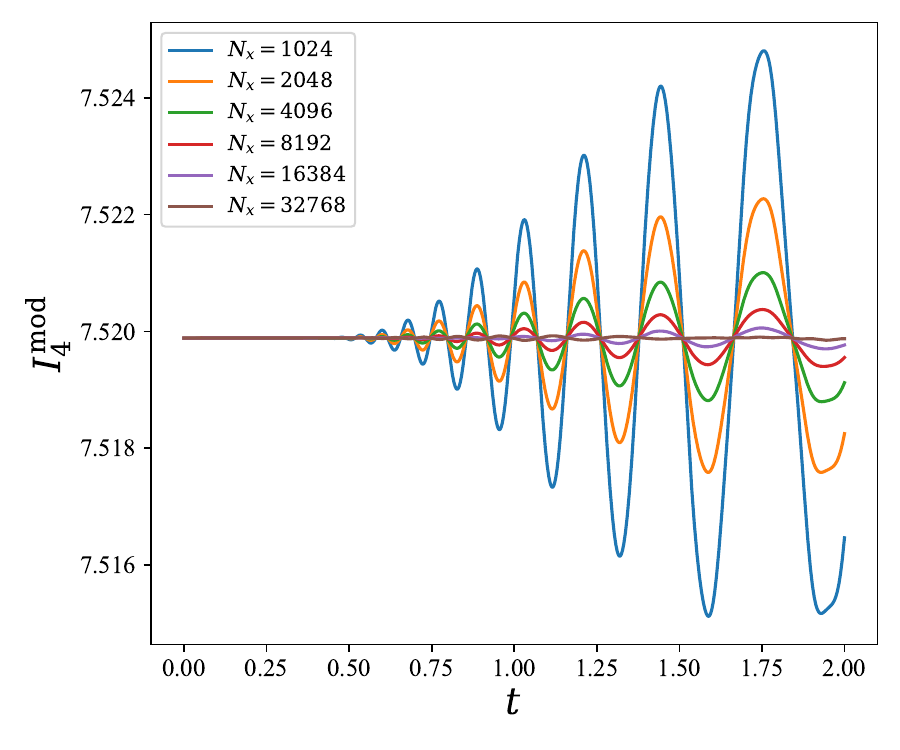}
\end{subfigure}
\vspace{-0.5cm}
\caption{
Effect of boundary-flux correction on the center-of-mass conservation law for the KdV equation with Gaussian initial data.
Left: Comparison of original $I_4$~\eqref{eq:kdv_I4_app} and $I_4^{\mathrm{mod}}$~\eqref{eq:kdv_I4_mod_app}.
Right: Spacial resolution dependence of $I_4^{\mathrm{mod}}$ (right).
}
\label{fig:kdv_resolution}
\end{figure}

Figure~\ref{fig:kdv_resolution} illustrates the effect of this modification for Gaussian initial conditions.
The corrected quantity $I_4^{\mathrm{mod}}$ substantially suppresses the drift observed in the original definition.
In contrast to the ZK equation, however, the accuracy of $I_4^{\mathrm{mod}}$ in the KdV case exhibits a stronger dependence on spatial resolution, since the correction relies on pointwise evaluation of boundary terms.
This sensitivity is purely numerical and can be readily mitigated by increasing the number of grid points, which is computationally inexpensive in one spatial dimension.

In summary, center-of-mass-type conserved quantities analogous to those of the ZK equation can also be defined for $(1+1)$-dimensional nonlinear dispersive systems such as the KdV equation.
The apparent violation of these conservation laws in finite-domain simulations arises from boundary-induced effects associated with dispersive radiation and can be systematically corrected using modified invariants.
This example reinforces the generality of the boundary-flux correction framework developed in the main text.

\end{appendix}


\vspace{0.5cm}

\section*{References}
\bibliography{zk_modified}


\end{document}